\documentclass[reprint,aps,amsmath, amssymb,groupedaddress]{revtex4-1}
\usepackage{amsmath}
\usepackage{amssymb}
\usepackage{amsfonts}
\usepackage{graphicx}
\usepackage{cases}
\usepackage{float}
\usepackage{dcolumn}
\usepackage{bm}

\begin{document}
\title{A machine learning based Bayesian optimization solution to nonlinear responses in dusty plasmas}
\author{Zhiyue Ding, Lorin S. Matthews and Truell W. Hyde}
\begin{abstract}
Nonlinear frequency response analysis is a widely used method for determining system dynamics in the presence of nonlinearities. In dusty plasmas, the plasma-grain interaction (e.g., grain charging fluctuations) can be characterized by a single particle nonlinear response analysis, while grain-grain nonlinear interactions can be determined by a multi-particle nonlinear response analysis. Here, a machine learning-based method to determine the equation of motion in the nonlinear response analysis for dust particles in plasmas is presented. Searching the parameter space in a Bayesian manner allows an efficient optimization of the parameters needed to match simulated nonlinear response curves to experimentally measured nonlinear response curves.
\end{abstract}

\maketitle
\section{Introduction}
Machine learning (or deep learning) has recently become one of the hottest analysis techniques in the scientific world as application of this powerful numerical method has proven useful in solving problems across a wide range of fields. For example, convolutional neural networks (LeNet, AlexNet) \cite{LeNet, AlexNet} now fulfill object recognition tasks to a high degree of accuracy, recurrent neural networks (LSTM) \cite{LSTM} are improving computational understanding of natural language, reinforcement learning agents are out-performing human experts in strategic decision making (AlphaGo) \cite{AlphaGo, AlphaGo2}, and Generative Adversarial Networks (GANs) \cite{GANs} are showing the ability to create music, paintings and dialogue in a human manner. In addition to industrial applications, machine learning techniques are now also being applied to solve physics problems. Examples include the prediction of molecular atomization energies by employing regression models \cite{Rupp2012}, the application of a neural network to solve quantum many-body problems \cite{Carleo2017}, and crystallization recognition through the use of a shallow neural network \cite{Dietz2017}. 

In this paper, a machine learning-based method is applied to nonlinear response problems in dusty plasmas. Dusty plasmas \cite{Melzer1996, Fortov2005, Morfill2009} are systems containing both weakly ionized gas and charged micron-sized dust particles. Due to the higher thermal velocities of electrons compared to ions, dust particles in a dusty plasma become negatively charged \cite{Goree1994} in response to the frequent collisions between the plasma particles and the dust grain's surface. Dust particle behavior in plasmas is determined by many factors, with the restoring confinement caused by the balance between the electrostatic force and gravity, the neutral gas drag, and particle-particle interactions between dust particles among the primary of these. Understanding the physics behind dust particle behavior (i.e., investigating these factors) is one of the most important tasks in dusty plasmas. One of the ways in which this can be fulfilled is by studying the response of the particles to external excitations \cite{Carstense2011, Ding2019}. This is known as the nonlinear frequency response analysis, which has a wide application in mechanics, material science and nano science \cite{PENG2007, Cottone2009, Samanta2015}.

Here, a Bayesian optimization framework \cite{Mokus1975} is used to resolve a nonlinear response analysis \cite{Ivlev2000, Zafiu2001, Wang2002, Ding2018} in a numerical manner in a dusty plasma. The undetermined coefficients in a dust grain's equation of motion will be derived by optimizing the simulated motion to match that obtained from experimental results which will be compared to the analytic results from a multiple-scale perturbation method. Also, we will measure the nonlinearity to the 4th order in displacement, which helps correctly characterize the potential energy of particle in the plasma sheath but has not been investigated before .

It is important to note that this framework is not limited to nonlinear response analysis, but can also be applied to the more general case of physics problems where the experimental results can be reproduced by simulations. In these cases, undetermined physics quantities can be revealed efficiently (especially when the simulation is very computational expensive) by optimizing the simulations to experimental results in this Bayesian manner. 

\section{Experiment and Bayesian optimization}
\begin{figure}[htbp]
	\centering
	\includegraphics[width=7cm,height=5cm]{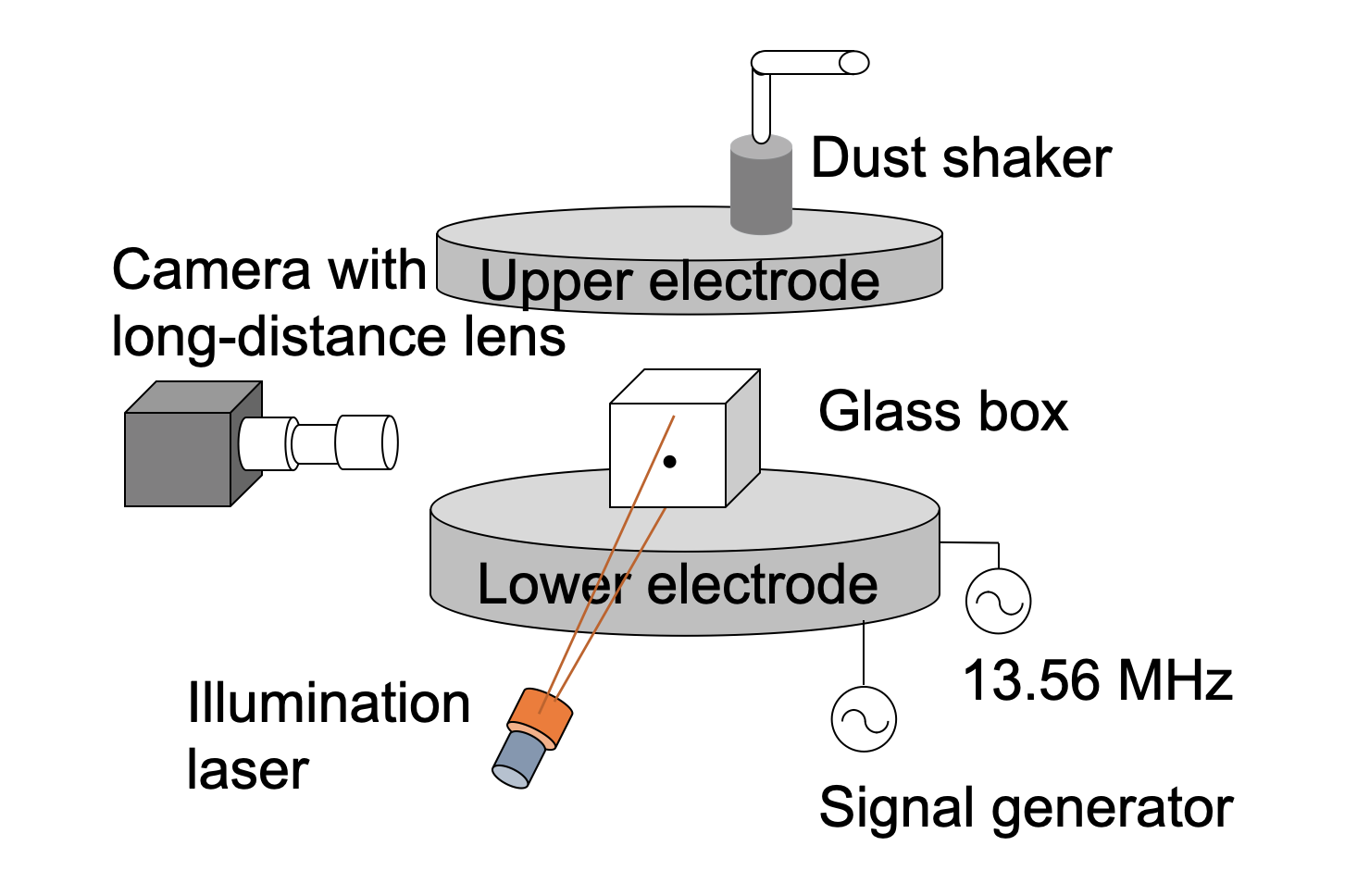}
	\caption{Sketch of the modified GEC RF reference cell.}
	\label{GEC}
\end{figure}

The experiment which will be discussed in this paper was conducted in a modified Gaseous Electronics Conference (GEC) RF reference cell (see Fig. \ref{GEC}) filled with argon gas. A single melamine formaldehyde (MF) particle having a diameter of 8.89 $\pm$ 0.09 $\mu m$ was inserted into a glass box (height: 20 mm, length: 18 mm, width: 18 mm) placed on the lower electrode which was powered at 13.56 MHz. The plasma power and pressure were fixed at 1.68 W and 40 mTorr, respectively. The MF particle was levitated in the plasma sheath region due to the balance between gravity and the electrostatic force produced by the negatively charged lower electrode. The dust particle was illuminated by a laser sheet (wavelength of 660 $nm$) with the resulting motion recorded by a high speed camera mounted at the side port of the cell at a rate of 500 fps. A primary amplitude-frequency response curve was measured by applying a sinusoidal excitation signal to the lower electrode with a fixed amplitude at various frequencies. Particle motion was recorded and then transformed into the frequency domain (FFT spectrum) using a Fourier transform at each value of the excitation frequency. The peak height of the FFT spectrum at the excitation frequency was measured, providing the primary response at this excitation. The secondary (super-harmonic) response to the excitation (a nonlinear response), can also be measured from the peak height of the FFT spectrum at twice the excitation frequency. 

The motion of a single such particle levitating inside the plasma sheath under a vertical sinusoidal excitation can be modeled as a forced oscillator \cite{Ivlev2000},
\begin{align}
&\ddot{x}+\mu \dot{x}+\omega^2x+\alpha x^2+\beta x^3=Fexp{(i\Omega t)}+c.c.,
\label{model}
\end{align}
where $\mu$ is the neutral drag coefficient, $\omega$ is the restoring constant, $\alpha$ and $\beta$ are the second and third order derivatives of the restoring field, $\Omega$ is the frequency of the sinusoidal excitation, $F$ is the amplitude of the excitation (in units of acceleration) and c.c. stands for the complex conjugate. Usually, the effective restoring force experienced by the particle at equilibrium can be approximated as a linear function in displacement (i.e., $-\omega^2x$ where $\omega$ is considered the natural resonance frequency) under the assumption that the particle is levitating in a region where the sheath can be considered to exhibit a perfect parabolic sheath potential \cite{Tomme2000, Carstensen2012}. Unfortunately, this approximation becomes invalid in most realistic situations, such as where charge fluctuations are considered \cite{Zafiu2001, Wang2002}, or when the oscillation of the dust particle is so large that the sheath potential can no longer be considered parabolic. In this case, it is necessary to extend the restoring force to the nonlinear regime as $-\omega^2x-\alpha x^2-\beta x^3$ with terms higher than $O(x^3)$ ignored for simplicity.

Based on Eq. \ref{model}, the particle motion $x(t)$ as a function of time under an excitation with frequency $\Omega$ can be simulated (or numerically solved) given a set of known parameters $\theta=\{\mu, \omega, \alpha, \beta, F\}$. In this case, the particle motion was simulated employing the velocity Verlet algorithm, which updates the position and velocity in each iteration as 
\begin{align}
\begin{split}
x(t+dt)=x(t)+v(t)dt+\frac{a(t)}{2}(dt)^2,\\
v(t+dt)=v(t)+\frac{a(t+dt)+a(t)}{2}dt,
\label{verlet}
\end{split}
\end{align}
where $dt$ is the time step of the simulation, $v(t)$ is the velocity at time $t$ and $a(t)$ is the acceleration normalized by the particle mass at time $t$ as determined by Eq. \ref{model}
\begin{align}
a(t)=-\mu v(t)-\omega^2x(t)-\alpha x^2(t)-\beta x^3(t)+2Fcos(\Omega t).
\label{accelation}
\end{align}

Following the same approach described above for measuring the amplitude-frequency response curves from experiment, the primary and secondary amplitude-frequency response curves can also be measured from the simulated particle motion $x(t)$ by varying the excitation frequency $\Omega$ over the range of excitain frequencies used in the experiment.

This allows a parameter set $\theta^*=\{\mu^*, \omega^*, a^*, b^*, F^*\}$ characterizing the properties of the dust motion (which depend on properties of the nearby plasma environment) to be determined by searching the parameter space for the optimal set of parameters that generates a simulated amplitude-frequency response curve resembling the experimentally measured amplitude-frequency response curve. This process can be quantified as
\begin{align}
\theta^*=\underset{\theta}{\operatorname{argmin}}(L(R_{e}, R_{s}(\theta))) ,
\label{abstract}
\end{align}
where $R_{e}$ represents the experimentally measured response curves and $R_{s}(\theta)$ represents the simulated response curves for a given set of parameters $\{\mu, \omega, \alpha, \beta, F\}$. $L(R_e, R_s)$ is a measure of the difference between the experimentally measured and simulated response curves. In order to quantify this difference, we define $L$ as a function $L: \theta=\{\mu, \omega, \alpha, \beta, F\}\mapsto \mathbb{R}$ that maps a set of parameters to a real value which measures the `distance' between the experimentally measured and the simulated response curve as
\begin{align}
L(\theta)=\sum_{i=1}^N (\frac{r_e(\Omega_i)-r_s(\Omega_i, \theta)}{r_e(\Omega_i)})^2,
\label{measure}
\end{align} 
where $r_e(\Omega_i)$ and $r_s(\Omega_i)$ are the experimentally measured and simulated response amplitudes at the excitation frequency $\Omega_i$, respectively. The summation is carried out over the excitation frequency span conducted in the experiment. The square operation is taken to ensure that $L(\theta)$ does not yield negative values, which guarantees the existence of minimal points. It is important to mention that this type of loss function $L(\theta)$ may not be well-defined everywhere. For unrealistic parameters sets, i.e., sets that either have no physical meaning or are not suitable for describing the condition of the plasma sheath, this simulation of nonlinear response curves diverges, resulting in an undefined distance function. In these cases, a large value is assigned to the distance function (e.g., $L=10^5$) in order to ensure optimization success. 

As can be seen from Eq. \ref{measure}, calculation of the loss function $L(\theta)$ for even one set of parameters requires multiple simulations of the particle's motion, i.e., one for each excitation frequency used for measuring the response curve in the experiment. For example, the response curve shown in Fig. \ref{p1_response} requires 71 independent simulations to calculate the distance function for just one set of parameters. This is computationally expensive and, as such, a minimization of the distance function $L(\theta)$ based on a random search of the parameter space $\theta$ is infeasible. 

Therefore, this loss function is minimized employing a Bayesian optimization. This technique has shown great promise in machine learning, especially for the fine tuning of neural networks for model selection. Now, instead of randomly searching the parameter space and then conducting simulations for each set, only those parameter sets selected in a Bayesian manner are simulated. A surrogate function $f$ is introduced to model the distribution of the value of the loss function $L(\theta)$. The posterior distribution of this surrogate function at the parameter $\theta$ given the data observed $\mathcal{D}_{1:t}=\{\theta_{1:t}, L(\theta_{1:t})\}$ can now be derived using the Bayes law
\begin{align}
p(f|\theta; \mathcal{D}_{1:t})=\frac{p(\theta|f; \mathcal{D}_{1:t})p(f; \mathcal{D}_{1:t})}{p(\theta; \mathcal{D}_{1:t})},
\label{surrogate_model}
\end{align} 
where Tree-structured Parzen density estimators \cite{Bergstra2011} (a generative model) are used to model the likelihood function $p(\theta|f; \mathcal{D}_{1:t})$ defined as 
\begin{equation}
p(\theta|f; \mathcal{D}_{1:t})=\begin{cases}
l(\theta), \text{if } f<f^*\\
g(\theta), \text{if } f\geq f^*.
\end{cases}
\label{likelihood}
\end{equation}
In this likelihood function, $l(\theta)$ and $g(\theta)$ are non-parametric Parzen density estimators (i.e., Gaussian mixtures) of the likelihood for the data to have a value of the loss function $L(\theta)$ smaller and greater than a threshold $f^*$, respectively. As such, the marginal distribution of the parameter set given the observed data set $\mathcal{D}_{1:t}$ (the denominator of Eq. \ref{surrogate_model}) can in turn be calculated as
\begin{align}
\begin{split}
p(\theta;\mathcal{D}_{1:t})=&\int_{-\infty}^{\infty}p(\theta|f;\mathcal{D}_{1:t})p(f;\mathcal{D}_{1:t})df\\
=&(l(\theta)-g(\theta))\int_{-\infty}^{f^*}p(f;\mathcal{D}_{1:t})df+g(\theta).
\end{split}
\label{marginal}
\end{align}

The criteria for exploring the overall parameter space is to choose the next set of simulation parameters that maximizes the expected improvement $\mathbb{E}[max(f^*-f,0)]$ \cite{Jones1998} as
\begin{align}
\begin{split}
\theta_{t+1} =&\underset{\theta}{\operatorname{argmax}}\int_{-\infty}^{\infty}\operatorname{max}(f^*-f, 0)p(f|\theta; \mathcal{D}_{1:t})df\\
=&\underset{\theta}{\operatorname{argmax}}\frac{\int_{-\infty}^{f^*}(f^*-f)p(f; \mathcal{D}_{1:t})df}{\frac{g(\theta)}{l(\theta)}(1-\int_{-\infty}^{f^*}p(f;\mathcal{D}_{1:t})df)+\int_{-\infty}^{f^*}p(f;\mathcal{D}_{1:t})df}\\
=&\underset{\theta}{\operatorname{argmax}}\frac{l(\theta)}{g(\theta)},
\end{split}
\label{next_parameter}
\end{align}
where the last equation holds since the cumulative distribution $\int_{-\infty}^{f^*}p(f;\mathcal{D}_{1:t})df$ is strictly less than 1. As shown, this result is not affected by the exact form of the prior $p(f; \mathcal{D}_{1:t})$. As such, the next parameter set whose distance function will be simulated is chosen to maximize the quotient of the Parzen density estimators $l(\theta)/g(\theta)$. As each new simulation is conducted, the data set $\mathcal{D}$ will be updated with the new simulated data points (e.g., the data set $\mathcal{D}_{1:t}$ is updated to $\mathcal{D}_{1:t+1}$ by adding a new simulated point $\{\theta_{t+1}, L(\theta_{t+1})\}$). The posterior distribution of the surrogate function Eq. \ref{surrogate_model} will also be updated accordingly, eventually resembling the behavior of the real distance function Eq. \ref{measure}. Notice that the threshold $f^*$ in Eqs. \ref{likelihood} and \ref{next_parameter} are the up-to-date optima (i.e., the minima of the loss function updated to the data $\mathcal{D}$). 

Since secondary responses (as nonlinear responses) are very sensitive to nonlinear terms, i.e., $\alpha x^2$ and $\beta x^3$, while the primary responses are more sensitive to the linear terms, it is necessary to minimize the loss functions for both primary and secondary responses simultaneously. One simple way of achieving this is to minimize a weighted sum of these two loss functions rather than minimizing them individually (e.g., $L=L_p+0.05L_s$). 

\section{Results}
\begin{figure}[htbp]
	\centering
	\includegraphics[width=9cm,height=4.2cm]{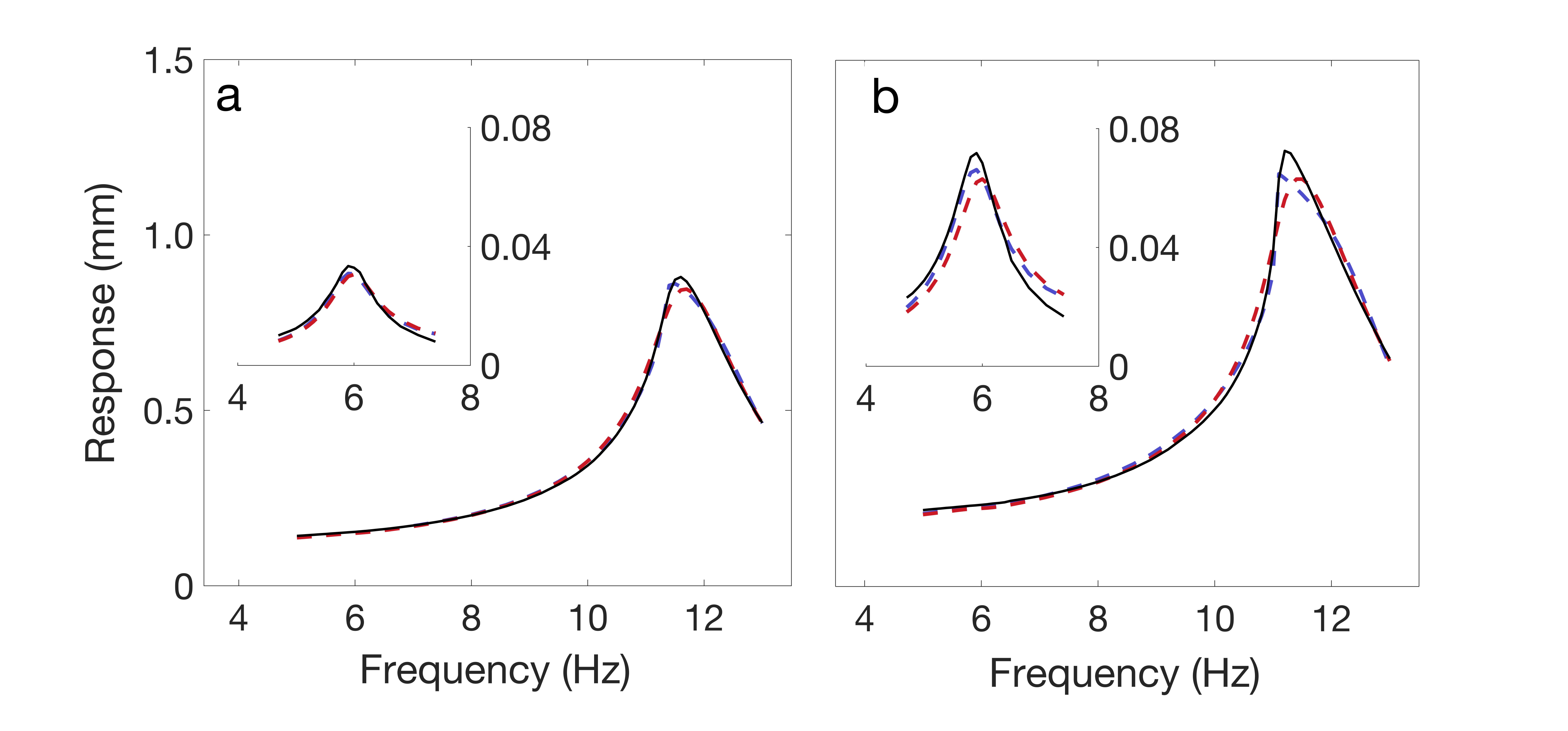}
	\caption{The primary Bayesian optimized (based on Eq. \ref{model}) and experimentally measured response curves plotted in dashed red and solid black, respectively, under a) 1.0 V excitation and b) 1.5 V excitation. The corresponding secondary (super-harmonic) response curves are shown in the subplots with the same color distribution. Dashed blue curves are the Bayesian optimized response curves based on Eq. \ref{model2}.}
	\label{p1_response}
\end{figure}

Figure. \ref{p1_response} shows the Bayesian-optimized simulated primary response curves (dashed red curves) of a single dust particle levitated in the plasma sheath in the GEC RF reference cell at a plasma power of 1.68 Watts and pressure of 40 mTorr. The corresponding secondary response curves are shown in the subplots. Particles excited under excitation amplitudes of 1.0 V and 1.5 V are plotted in Fig. \ref{p1_response}a and Fig. \ref{p1_response}b, respectively. As shown, the optimized response curves (dashed red curves simulated according to Eq. \ref{model}) resemble the experimentally measured responses curves (solid black curves) in both the primary and secondary regions. Also, note that the spring softening phenomenon (i.e., the nonlinear phenomenon that results in the primary resonance peak being `bent' in the low frequency direction) becomes more obvious as the excitation amplitude increases (Fig. \ref{p1_response}a).

The corresponding optimized parameters obtained from model 1 (Eq. \ref{model}) are calculated as the average over five independent trials of the optimizing experiment and their values are listed in Table \ref{tab_parameters1} (method `Model 1'), with the corresponding Coefficient of Variations (CV) shown in parentheses. Notice that the sign of the coefficient of the quadratic nonlinearity $\alpha$ is irrelvant in this response analysis since it only changes the direction of the asymmetric motion of the dust particle. Even though there is a large variation in randomness in this Bayesian search of the parameter space, the optimizing experiment converges to yield consistent results as evidenced by the low CV. The relatively high CV for the parameter $\beta$ (the coefficient of the cubic nonlinearity) is due to the fact that the response curves are robust in responding to variation of nonlinearities of higher order. As such, a small variation in $\beta$ will not significantly perturb the entire response curve. 

\begin{figure}[htbp]
	\centering
	\includegraphics[width=9cm,height=4.2cm]{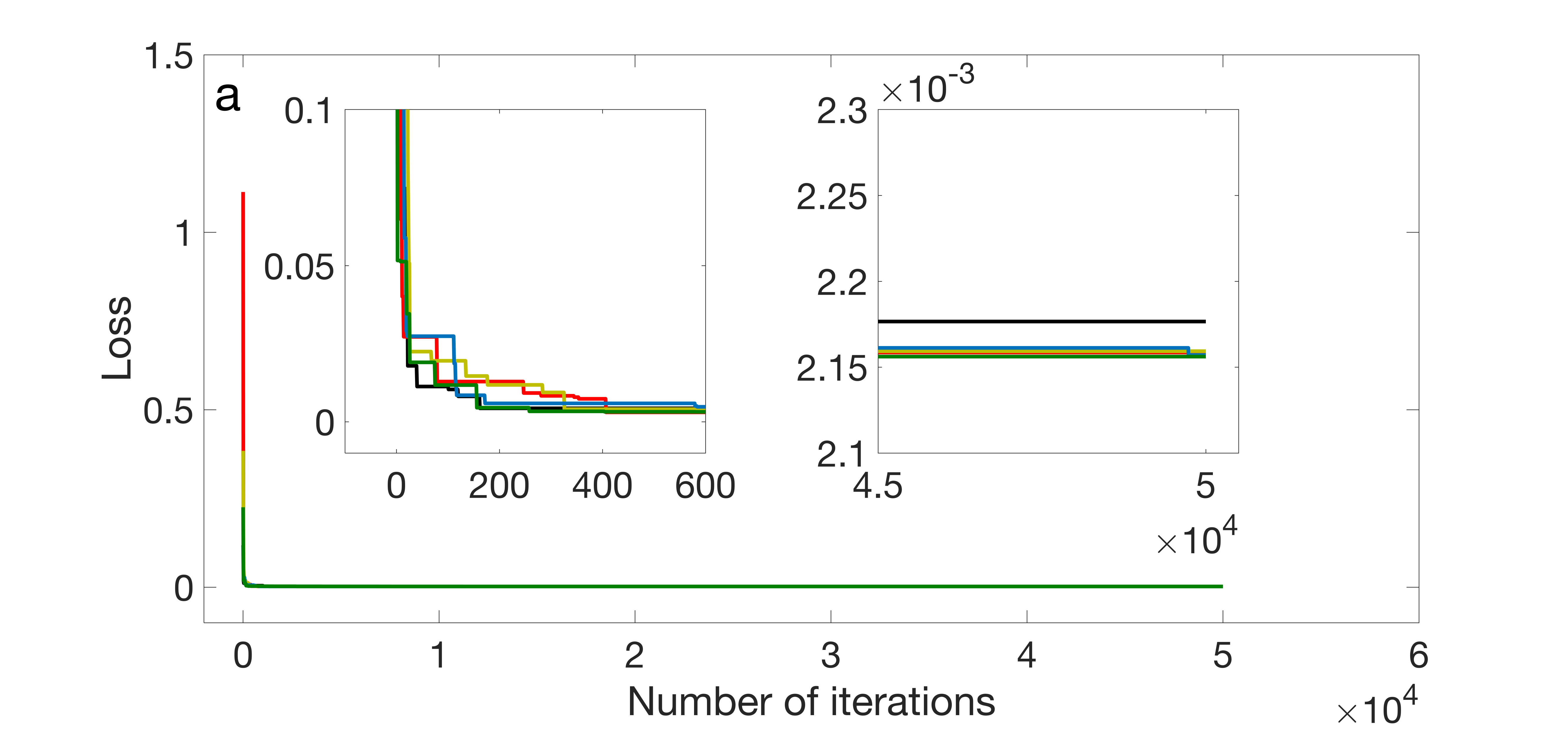}
	\centering
	\includegraphics[width=9cm,height=4.2cm]{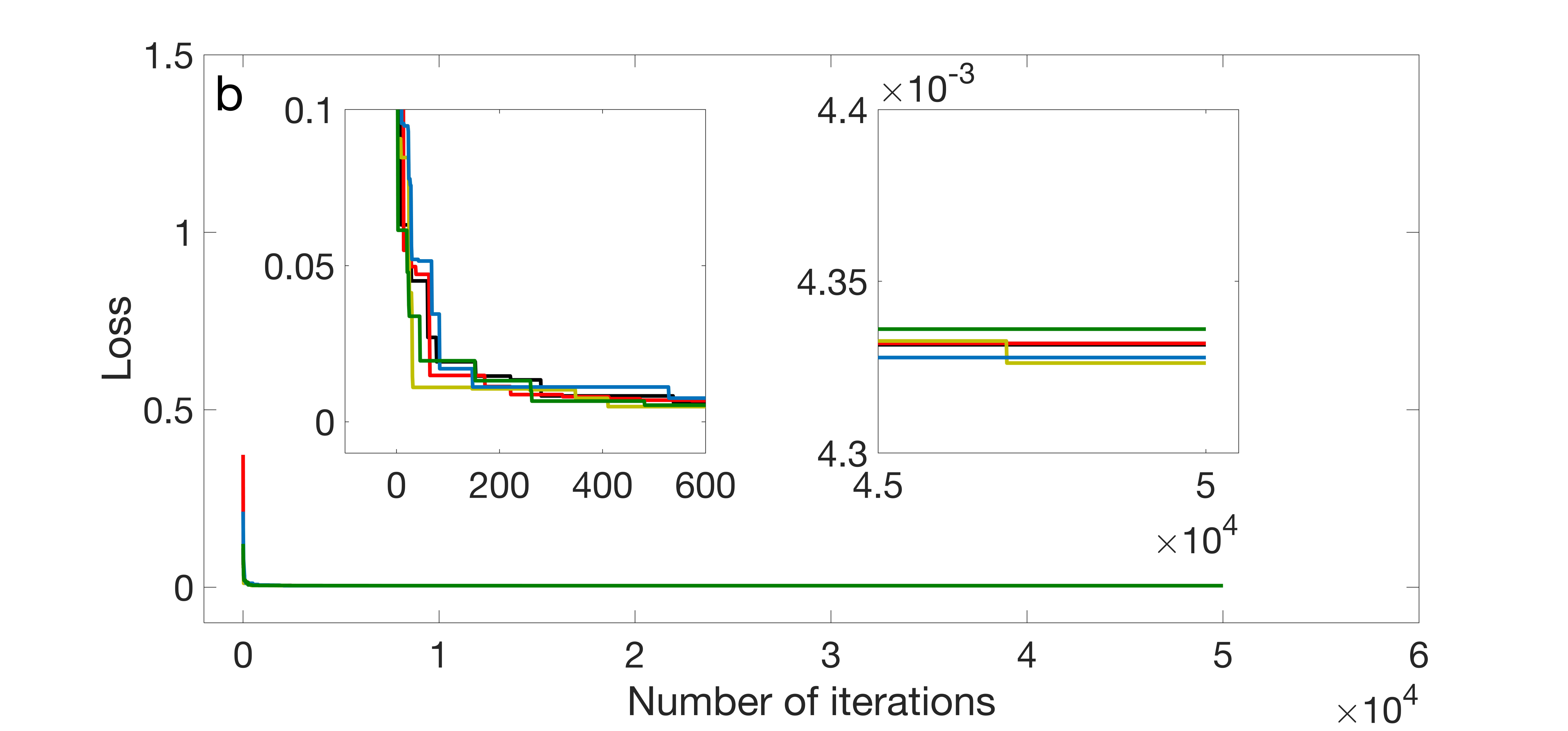}
	\caption{The loss (the `distance' between the experimentally measured and the simulated response curve) as a function of the number of iterations for a dust particle excited under an excitation amplitude of a) 1.0 V and b) 1.5 V. Colors denote the five independent trials.}
	\label{loss}
\end{figure}

Fig. \ref{loss} shows the loss (Eq. \ref{measure}) as a function of the number of iterations for the dust particle excited at both 1.0 V (a) and 1.5 V (b), each of which has five independent experimental trials. As shown, the loss values decrease rapidly after the first several iterations, reaching convergence after a few hundred iterations. (This again indicates the efficiency of the presented Bayesian optimization method in exploitation of the parameter space.) However, in order to boost overall accuracy and ensure wider exploration of the parameter space, we conducted a large number of iterations. The observed higher convergency loss value for the dust particle under 1.5 V excitation ($\approx 4.3\times 10^{-3}$) as compared to that under 1.0 V excitation ($\approx 2.2\times 10^{-3}$) can be attributed to the increased difficulty of capturing the spring softening phenomenon as the excitation amplitude becomes larger. (Compare Fig. \ref{p1_response}a and Fig. \ref{p1_response}b.)
\begin{widetext}

\begin{table}[h]
\centering
\renewcommand*{\arraystretch}{1.8}
\setlength{\tabcolsep}{1mm}

\begin{tabular}{ccccccc}
 \hline
 \hline

Methods & $\mu$ ($s^{-1}$) & $\omega$ (Hz) & $|\alpha|$ ($\mu m^{-1}\cdot s^{-2}$) & $\beta$ ($\mu m^{-2}\cdot s^{-2}$) & $F$ ($\mu m^{-1}\cdot s^{-2}$) & $|\gamma|$ ($\mu m^{-3}\cdot s^{-2}$)\\
 \hline
Excitation 1.0 V: \\
Model 1 &10.1 (0.6\%)& 11.9 (0.1\%) & 2.1 (0.9\%)& $3.9\times 10^{-4}$ (5.7\%) & $6.3\times 10^5$ (0.4\%)\\
Multiple-scale &  9.6& 11.9& 2.0& $3.5\times 10^{-4}$ & $6.2\times 10^5$&\\
Model 2 &$9.7$ (1.5\%)& $11.8$ (0.3\%)& $2.0$ (1.2\%)& $14.2\times 10^{-4}$ (7.3\%)&$6.2\times 10^5$ (0.4\%) & $1.8\times 10^{-6}$ (6.6\%)\\
 \hline
Excitation 1.5 V: \\
Model 1 & $10.9$ (0.1\%)& $11.9$ (0.0\%) & $2.1$ (0.8\%) & $3.8\times 10^{-4}$ (4.0\%) & $9.4\times 10^5$ (0.2\%)&\\
Multiple-scale & $10.2$ & $11.9$ &$1.9$ & $2.1\times 10^{-4}$ & $9.1\times 10^5$\\
Model 2 & $10.3$ (1.3\%)& $11.7$ (0.2\%)&$2.0$ (1.2\%)& $15.5\times 10^{-4}$ (3.5\%)&$9.0\times 10^5$ (0.6\%)&$1.2\times 10^{-6}$ (5.6\%)\\
\hline
\hline
\end{tabular}
\caption{The parameter space measured for Model 1 (Eq. \ref{model}) from the Bayesian optimization method and the multiple-scale perturbation method, and for Model 2 (Eq. \ref{model2}) from the Bayesian optimization method are shown in this table. For the Bayesian optimization method, the measurments are averages of five independent experimental trials, with the corresponding coefficients of variance shown in parentheses.}
\label{tab_parameters1}
\end{table}

\end{widetext}

\section{Multiple-scale perturbation method}
The parameters can also be derived analytically by solving the equation of motion (Eq. \ref{model}) employing the multiple-scale perturbation method. The details of this method are given in refs \cite{Nayfeh1979, Ding2018}, with the main results needed for the analysis given below.

Assuming an external excitation at a frequency of approximately half that of the oscillator resonance frequency $\omega$, i.e., $\Omega\approx\frac{1}{2}\omega$, the solution to Eq. \ref{model}, to first order, yields
\begin{align}
\begin{split}
x(t) =&\frac{F}{\omega^2-\Omega^2}cos(\Omega t)\\
&-\frac{\alpha F^2}{4\omega (\omega^2-\Omega^2)^2\sqrt{\frac{\mu^2}{4}+(2\Omega-\omega)^2}}sin(2\Omega t - \phi),
\end{split}
\label{superharmonic}
\end{align}
where $\phi$ is the shifted phase which is dependent on the excitation frequency as $\phi = arctan(\frac{4\Omega-2\omega}{\mu})$. The parameters $\omega$, $\mu$ and $\alpha$ are determined using the experimentally measured secondary response curve fitted to the steady state theoretical secondary response $\alpha F^2/4\omega (\omega^2-\Omega^2)^2\sqrt{\frac{\mu^2}{4}+(2\Omega-\omega)^2}$. Considering an external excitation having a frequency approximately equal to the oscillator resonant frequency, i.e., $\Omega\approx\omega$, the solution to Eq. \ref{model}, to first order of approximation yields
\begin{align}
\begin{split}
x(t) =A(\Omega)cos(\Omega t - \phi'),
\end{split}
\label{superharmonic}
\end{align}
where the shifted phase is $\phi'=\Omega-\omega-\beta$. By eliminating the secular term appearing in the equation of motion to second order of approximation, the steady state theoretical primary response $A(\Omega)$ can now be derived as
\begin{align}
\begin{split}
\frac{F^2}{4\omega^2}=(\frac{A\mu}{2})^2+[(\frac{9\beta\omega^2-10\alpha^2}{24\omega^3})A^3-(\Omega-\omega)A]^2.
\end{split}
\label{secular}
\end{align}

By fitting the corresponding experimentally measured primary response curve to Eq. \ref{secular}, the parameter space for the cubic nonlinearities $\beta$ and $F$ can now be investigated. The parameters obtained in this way are shown in Table \ref{tab_parameters1} (method `Multiple-scale'). As shown, the parameters measured from the Bayesian optimization are consistent with those measured from the multiple-scale perturbation, except for the value of $\beta$ under 1.5 V excitation (with approximately 57.6$\%$ percent difference). Notice that the measurements of the parameters from the multiple-scale perturbation serve as a benchmark and should not be considered as true values since they are derived and are precise only to the first order of approximation.

Due to limitations of the multiple-scale perturbation method, any extension of the model (Eq. \ref{model}) (e.g., including higher order nonlinearities of either displacement $x$ or velocity $\dot{x}$) and derivation of the corresponding approximate solutions would be tediously complicated. However, the Bayesian optimization scheme described here allows this process to be simplified greatly. As an example, the model is extended to include an additional nonlinearity of higher order in displacement $x$,
\begin{align}
&\ddot{x}+\mu \dot{x}+\omega^2x+\alpha x^2+\beta x^3+\gamma x^4=Fexp{(i\Omega t)}+c.c..
\label{model2}
\end{align}
Applying the Bayesian method, the optimized parameters are measured, with the results shown in Table \ref{tab_parameters1} (with method `Model 2') and the corresponding simulated response curves shown in Fig. \ref{p1_response} (dashed blue curves). Again, the corresponding CV for the five independent trials are shown in parentheses.

\section{Discussion and Conclusion}
\begin{figure}[htbp]
	\centering
	\includegraphics[width=7.83cm,height=3cm]{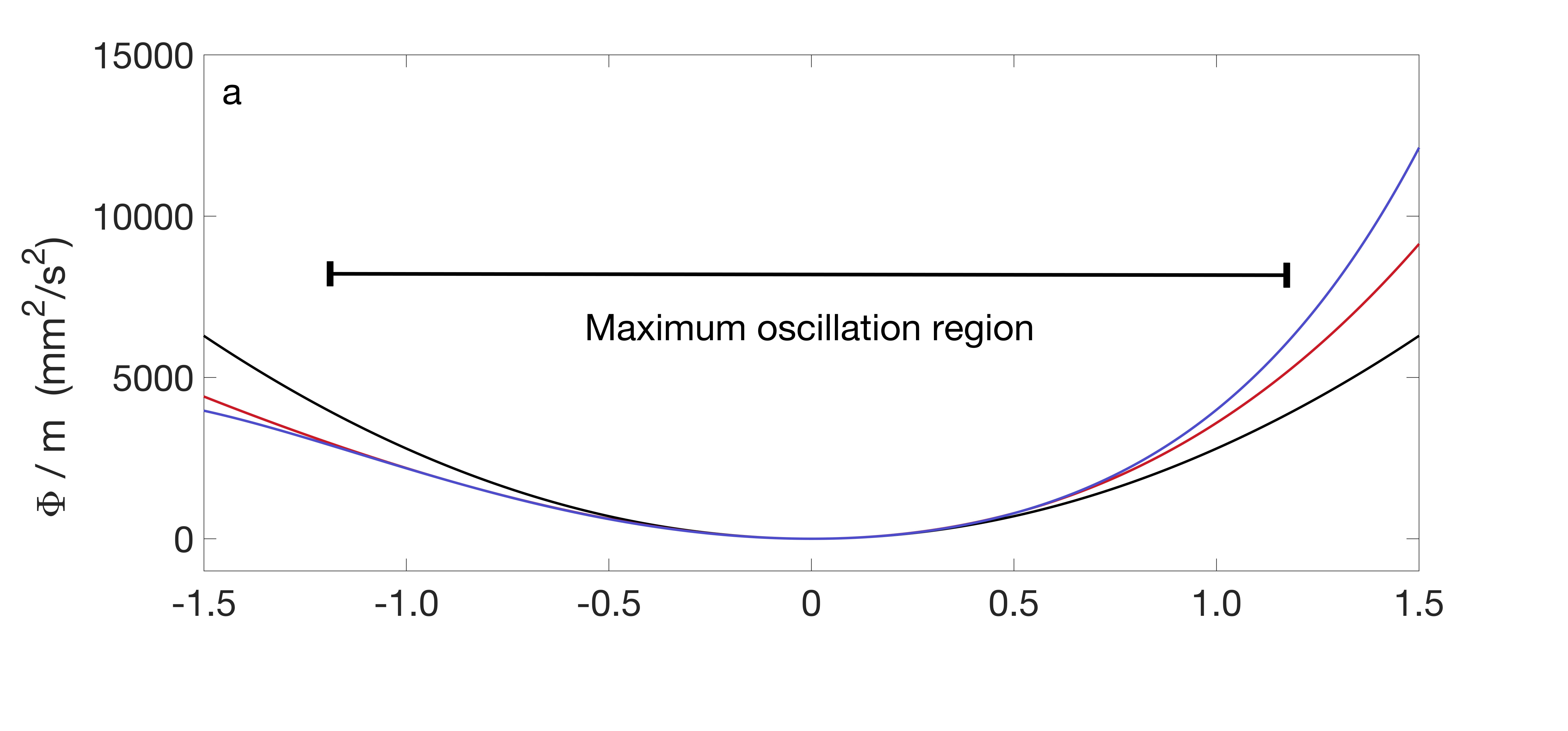}
	\centering
	\includegraphics[width=7.83cm,height=3cm]{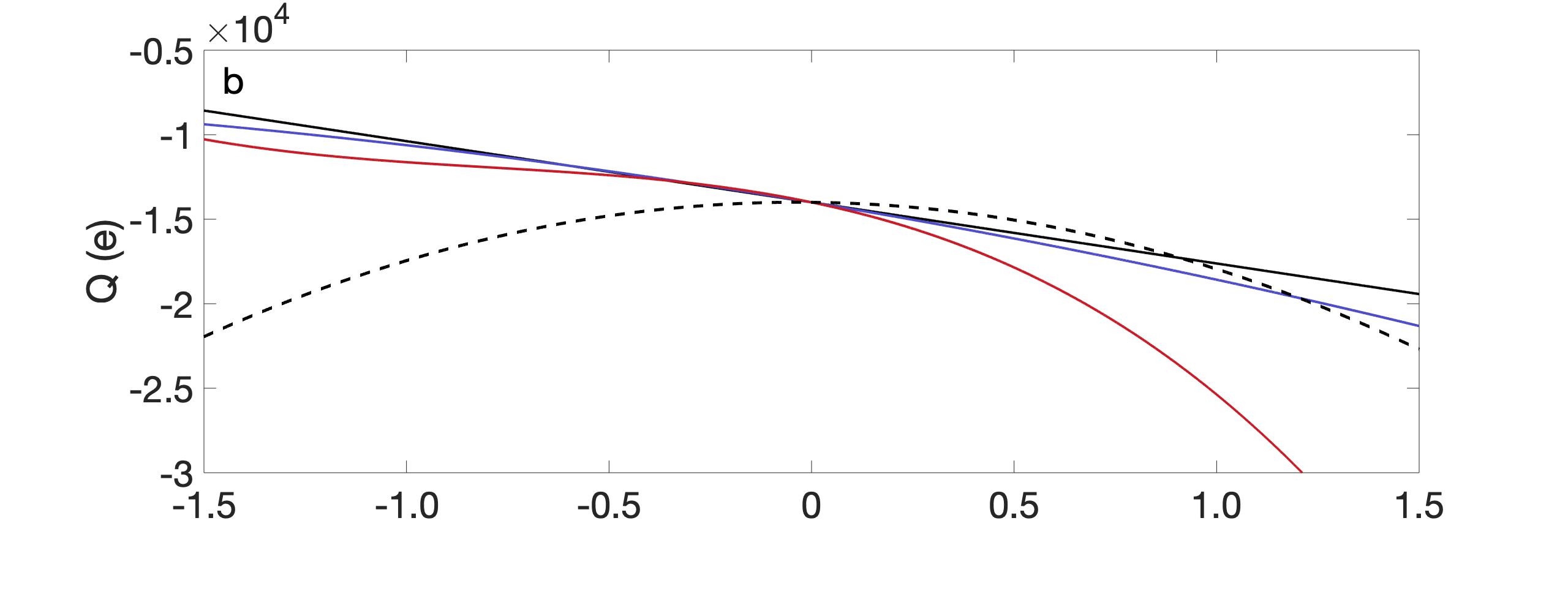}
	\centering
	\includegraphics[width=7.83cm,height=3cm]{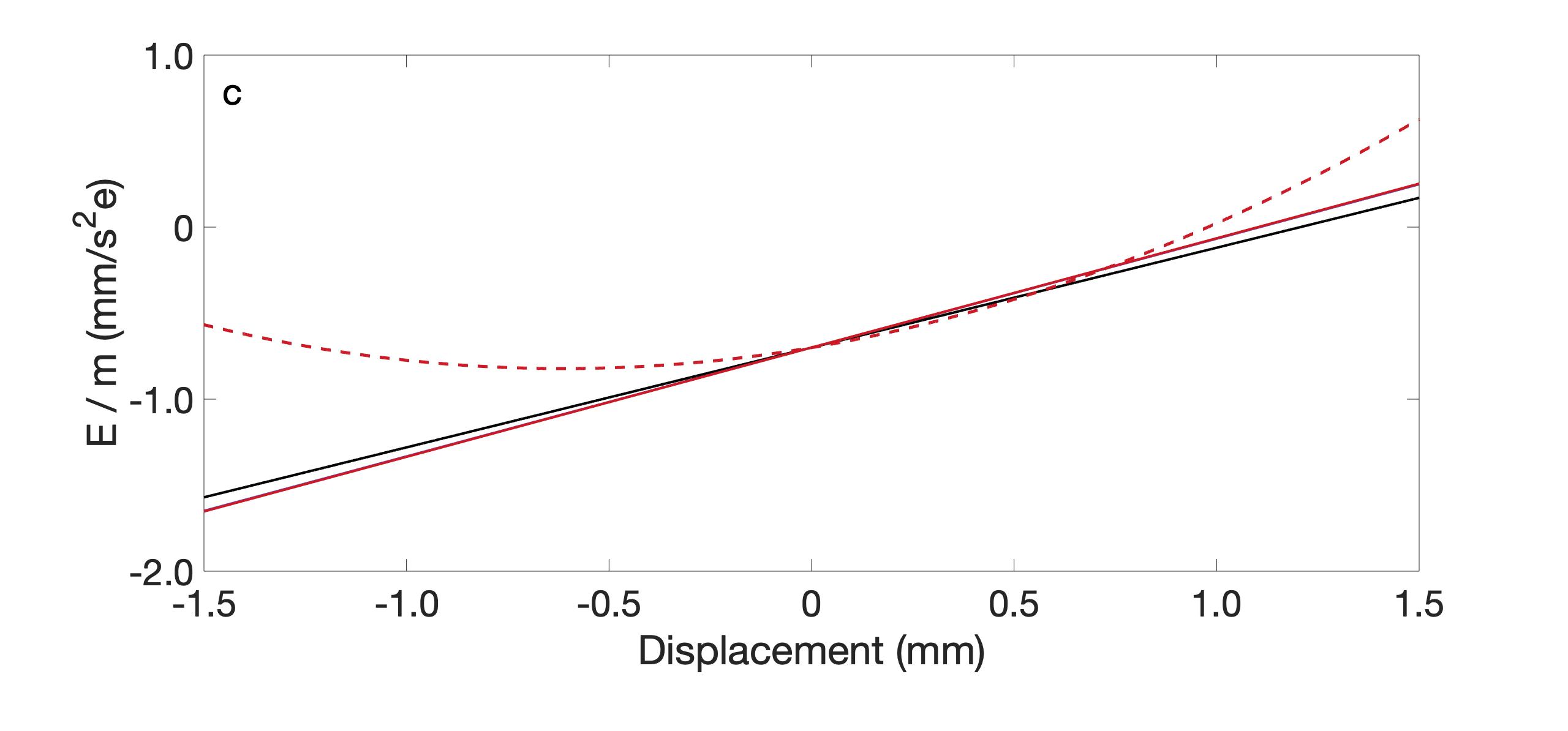}
	\caption{a) Restoring potential energy around the equilibrium levitation position (divided by the mass of the particle) for Model 1 (red) and Model 2 (blue). The perfect parabolic approximation is shown by the black line as a benchmark. The region over which a particle oscillates in this experiment (with maximum ampllitude of 1.2 $mm$) is also indicated.  Grain charge (b) and the electric field (c) in the vicinity of the equilibrium position. Red, blue and black solid lines correspond to the third order, second order and linear polymonial expansion of the grain charge with a linear electric field, respectively. The dashed lines correspond to the second order polynomial expansion for both grain charge and the electric field.}
	\label{potential_energy}
\end{figure}

By considering nonlinearities to fourth order, the primary response curves (dashed blue line) in Fig. \ref{p1_response}a more closely resemble the spring softening behavior than do those considering nonlinearities to third order (dashed red line). Also, the loss is further reduced, reaching $1.6 \times 10^{-3}$ for a 1.0-V excitation and $2.3 \times 10^{-3}$ for a 1.5-V excitation, as compared to values based on the model with third order nonlinearities (Eq. \ref{model}) which results in loss values of $2.2 \times 10^{-3}$ and $4.3 \times 10^{-3}$, respectively. This indicates a closer match of the simulated response curves to the experimentally measured ones. After introducing nonlinearities to the fourth order, the measured drag coefficient $\mu$, excitation amplitude $F$,  and the coefficient of the quadratic nonlinearity $\alpha$ more closely approach the values measured from the multiple-scale perturbation. However, it is noted that the coefficient for the cubic nonlinearities, $\beta$ exhibits a large deviation. Considering the conditions for the existence of the spring softening effect (seen from Eq. \ref{secular})
\begin{align}
\begin{split}
9\beta\omega^2-10\alpha^2<0,
\end{split}
\label{spring_softening}
\end{align}
the critical value of $\beta$ for the existence of the spring softening phenomenon can be derived as $\beta<\beta_c \approx8.1 \times 10^{-4}$ (by taking into consideration the fact that the measured values for the coefficient of the quadratic nonlinearity $\alpha$ are consistent in both models given by Eq. \ref{model} and Eq. \ref{model2}). In this case, a large value of $\beta$ as measured for both a 1.0-V excitation and a 1.5-V excitation, based on the model represented by Eq. \ref{model2}, seems to violate the condition of the existence of the spring softening phenomenon (Eq. \ref{spring_softening}). However, Eq. \ref{spring_softening} is derived from only the first order of approximation in the multiple-scale perturbation. Thus, although a response curve simulated with a parameter $\beta$ violating the condition given by Eq. \ref{spring_softening} still reveals the spring softening phenomenon, this indicates a limited ability and accuracy of the multiple-scale perturbation method to explain nonlinear responses since it ignores higher order nonlinearities. In order to accurately determine the coefficient of the cubic nonlinearities $\beta$ (an important factor characterizing the nonlinearity of the plasma sheath \cite{Ivlev2000, Zafiu2001}), the effects from higher order nonlinearities are important and should not be ignored. Fig. \ref{potential_energy}a shows the effective restoring potential energy $\Phi$ of the particle (divided by the particle mass) in the vicinity of its equilibrium position for both Model 1 (red) and Model 2 (blue). The difference in restoring potential energy is observable for these two models.

With the coefficient for the fourth order nonlinearity available, the change in the electric field and the grain charge at varying levitation positions can be further investigated. By considering an expansion in the electric field $E$ and grain charge $Q$
 \begin{align}
 \begin{split}
&E=E_0+E_1x+E_2x^2+E_3x^3,\\
&Q=Q_0+Q_1x+Q_2x^2+Q_3x^3,
\label{electro_force}
\end{split}
\end{align}
the electrostatic force can be written as
 \begin{align}
 \begin{split}
F_{stat}=&(E_0Q_0)+(E_0Q_1+E_1Q_0)x\\
&+(E_0Q_2+E_1Q_1+E_2Q_0)x^2\\
&+(E_0Q_3+E_1Q2+E_2Q_1+E_3Q_0)x^3\\
&+(E_1Q_3+E_2Q_2+E_3Q_1)x^4.
\label{electro_force}
\end{split}
\end{align}
The coefficents of the polynomial in this expansion are related to the corresponding coefficents in Eq. \ref{model2}. By assuming a linear electric field (i.e., $E_2=E_3=0$), the nonlinearities in charge $Q$ can be explored to the third order ($Q_3$) with $\gamma$ provided by the Bayesian optimzation approach (Model 2), while without $\gamma$, the nonlinearities in charge can only be explored up to the second order ($Q_2$). Fig. \ref{potential_energy}b shows the grain charge in the vicinity of the equilibrium position for the third order polymonial ($Q=Q_0+Q_1x+Q_2x^2+Q_3x^3$) and second order polymonial ($Q=Q_0+Q_1x+Q_2x^2$) expansion in red and blue, respectively. As a reference, a linear charge model ($Q=Q_0+Q_1x$) is also shown in black. The corresponding linear electric field (divided by the particle mass) for each charge expansion are shown in Fig. \ref{potential_energy}c. The equilibrium charge $Q_0\approx1.4\times 10^4 e$ was estimated by using the levitation position comparison method \cite{Ding2016}. In this method, a vertically aligned two particle pair is formed, and the difference in the levitation position for the upstream particle with and without the presence of the downstream particle (where the downstream particle is knocked out of the system using a laser pulse), is measured. As shown, the third order polymonial charge model predicts a weaker charge reduction in the downstream region, but a stronger charge accumulation in the region above the equilibrium position. 

Beyond assuming a linear electric field, we can also investigate the nonlinear expansions for the E-field and the grain charge simultaneously (i.e., $E=E_0+E_1x+E_2x^2$ and $Q=Q_0+Q_1x+Q_2x^2$). However, due to a lack of constraints, this investigation can only be explored to the second order for both electric field and grain charge even with $\gamma$ known. Fig. \ref{potential_energy}b and Fig. \ref{potential_energy}c show the result of this charge model and the corresponding nonlinear electric field as dashed lines. As shown, both the grain charge and the electric field can be very different from the cases where the electric field is assumed to be linear in the downstream region. This indicates a resonable assumption of a linear electric field in a close vicinity of the equilibrium position. It is instructive to compare these results against the usual linear models for both the particle charge and electric field. With the assumption of a linear electric field, the charge varies considerably from the linear charge model in the upstream direction. Conversely, with the assumption of a quadratic electric field, it is seen that the charge varies significantly from the linear charge model in the downstream direction. Future experiments may be designed to determine which of these models is correct.

In conclusion, a nonlinear response analysis for dust particles in plasma was provided employing a machine learning based method. An efficient technique for optimizing the comparison between numerically simulated and experimentally measured response curves by searching the parameter space in a Bayesian manner was described. Using this approach, the physical parameters characterizing the plasma conditions can be derived. The nonlinearity of the response was determined to the fourth order, which is necessary in order to accurately determine the the coefficients for lower-order nonlinearities, as well as to correctly characterize the potential energy of the particle in the sheath. Beyond the field of dusty plasmas, the proposed framework provides a general method for measuring physical quantities by optimizing simulation parameters to match experimental observations in an efficient manner, especially when the simulation is computationally expensive. 

\section{Acknowledgements}
Support from NASA grant number 1571701, NSF grant number 1740203 and NSF grant number 1707215 is gratefully acknowledged.

\section*{References}
\bibliography{paperNotes}

\end{document}